\documentclass[prl,showpacs,twocolumn,amsmath,amssymb,groupadress,superscriptaddress]{revtex4}
\usepackage{graphicx}
\usepackage{dcolumn}
\usepackage{bm}
\begin{document}

\preprint{APS/PRL}

\title{Structural anomalies at the magnetic transition in centrosymmetric BiMnO$_3$}

\author{E. Montanari}
\affiliation{Dipartimento di Chimica GIAF, Universit\`{a} di Parma, Parco Area delle Scienze 17A, 43100 Parma, Italy}
\author{G. Calestani}
\affiliation{Dipartimento di Chimica GIAF, Universit\`{a} di Parma, Parco Area delle Scienze 17A, 43100 Parma, Italy}
\affiliation{CNR-IMEM, Parco Area delle Scienze 37/A - 43010 Loc. Fontanini-Parma, Italy}\author{L. Righi}
\affiliation{Dipartimento di Chimica GIAF, Universit\`{a} di Parma, Parco Area delle Scienze 17A, 43100 Parma, Italy}
\author{E. Gilioli}
\author{F. Bolzoni}
\affiliation{CNR-IMEM, Parco Area delle Scienze 37/A - 43010 Loc. Fontanini-Parma, Italy}
\author{K.S. Knight}
\affiliation{ISIS facility, Rutherford
Appleton Laboratory-CCLRC, Chilton,
Didcot, Oxfordshire, OX11 0QX, United
Kingdom}
\author{P.G. Radaelli}
\affiliation{ISIS facility, Rutherford
Appleton Laboratory-CCLRC, Chilton,
Didcot, Oxfordshire, OX11 0QX, United
Kingdom}
\affiliation{Department of Physics and Astronomy, University College London, Gower Street, London WC1E 6BT, United Kingdom}
\date{\today}

\begin{abstract}
The structural properties of BiMn$O_3$  were determined from neutron powder diffraction data as a function of temperature and magnetic field. The structure at all temperatures was found to be centrosymmetric with space group $C2/c$, which is incompatible with ferroelectricity.  At T$_C\simeq 100 K$, we observed the onset of a large magnetoelastic strain, proportional to the square of the magnetization.  We interpret this structural rearrangement, together with the previously observed magnetodielectric anomalies, as due to the need to optimize the partially frustrated magnetic interactions.
\end{abstract}

\pacs{75.25.+z, 75.30.Kz, 75.80.+q,   77.80.-e, 77.80.Bh,  }
\maketitle

Multiferroic materials, for which two or all three of ferroelectricity, (anti-) ferromagnetism and ferroelasticity are observed in the same phase, have received renewal interest in recent years. Such systems are rare in nature but are potentially interesting for a wide array of technological applications \cite{fiebig, Ramesh}. In particular, BiMnO$_3$ has been extensively studied in both bulk and thin film forms as a candidate ferromagnetic (FM) ferroelectric (FE) material. BiMnO$_3$ shows FM ordering at the Curie temperature T$_C$ = 99-105 K, a unique feature among the Bi\textit{B}O$_3$ transition-metal series (\textit{B} = Cr-Ni). The observed ferromagnetism is related to ordering of the partial filled 3d$_{z^2}$ and 3d$_{x^2-y^2}$ orbitals on Mn, as evidenced by a cooperative Jahn-Teller distortion taking place well above T$_C$ \cite{Atou}. Dos Santos \textit{et al.} showed that BiMnO$_3$ is monoclinic above and below T$_C$, and proposed the non-centrosymmetry space group $C2$, which in principle allows ferroelectricity along the $b$ axis \cite{dos_Santos_1}. However, establishing experimentally that BiMnO$_3$ is also FE has proved considerably more difficult, due to the lack of single crystals and the low resistivity of BiMnO$_3$ films.   Band structure calculations performed in $C2$ by Hill and coworkers \cite{Hill_2} predicted the existence of ferroelectricity in BiMnO$_3$, and Shishidou \textit{et al.} later calculated a polarization of 0.52  C/cm$^2$ \cite{Shishidou}. dos Santos \textit{et al.} \cite{dos_Santos_2} showed FE hysteresis loops at room temperature and below, on both bulk and thin films samples, with a FE polarization of 0.043  C/cm$^2$ at 200 K, one order of magnitude lower than predicted.  Also, Sharan \textit{et al.} reported evidence for second-harmonic generation (SHG), strongly enhanced by applied electric field, consistent with the presence of FE domains in their thin-film samples\cite{Sharan_2}. All these data would indicate that BiMnO$_3$ is a \emph{proper} FE, since ferroelectricity is observed already at room temperature.  This rules out any mechanism based on symmetry breaking at the onset of magnetic ordering, as recently proposed, for example, for TbMnO$_3$\cite{Kimura2}. In support of  these results, a separate set of measurements shows quite convincingly a very strong magneto-electric coupling in BiMnO$_3$ near T$_C$. Kimura \textit{et al.} \cite{Kimura} investigated the magnetodielectric effect in BiMnO$_3$ bulk samples: they observed a pronounced anomaly in the dielectric constant $\epsilon$, and a very strong magnetic field dependence of $\epsilon$ near T$_C$, which they ascribe to the coupling between FE and FM order.

This substantial body of work has been very recently challenged by Belik \textit{et al.}\cite{Belik} After a detailed analysis of convergent beam electron diffraction and neutron diffraction data on bulk samples, Belik concluded that there is no evidence for inversion symmetry breaking in BiMnO$_3$, and proposed the \emph{centrosymmetric} space group $C2/c$, which cannot support ferroelectricity.  It is noteworthy that first-principle calculations on BiMnO$_3$ also show that the centrosymmetric $C2/c$ structure is more stable than the $C2$ structure at 0 K \cite{Shishidou_APS}. The simplest explanation for this clear contradiction between these data and previous literature has to do with sample differences, particularly between films and bulk.  In particular, the lower T$_C$ of thin films with respect to the bulk may indicate partial oxidation, leading to different polymorphs \cite{Montanari_1}.  Thin films are also affected by the presence of the substrate, which could reduce their symmetry.  However, Kimura's data \cite{Kimura} are impossible to dismiss on this ground, since they were collected on high-quality polycrystalline sample.  Nevertheless, the observation of a large magnetodielectric effect does not require \textit{ipso facto} the breaking of inversion symmetry, and may arise from structural anomalies of a different kind.  Here, we present neutron diffraction measurements collected on BiMnO$_3$ as a function of temperature (10 $\leq$ T $\leq$ 295 K)  and magnetic field (0 $\leq$ H $\leq$ 10 T).  In agreement with Belik \textit{et al}\cite{Belik}, we find that the \textit{centrosymmetric} space group $C2/c$ is perfectly adequate to model our data at all temperatures and fields.  We observed very significant structural anomalies upon entering the magnetically ordered phase. On the basis of the Landau theory of phase transitions, we propose that the large magnetodielectric effect observed by Kimura is related to the anomalies we observe, most likely through a subtle change in the environment of the highly polarizable Bi ion, and both are due to the need to optimize the partially frustrated magnetic interactions.

\begin{figure}
\includegraphics[scale=0.42]{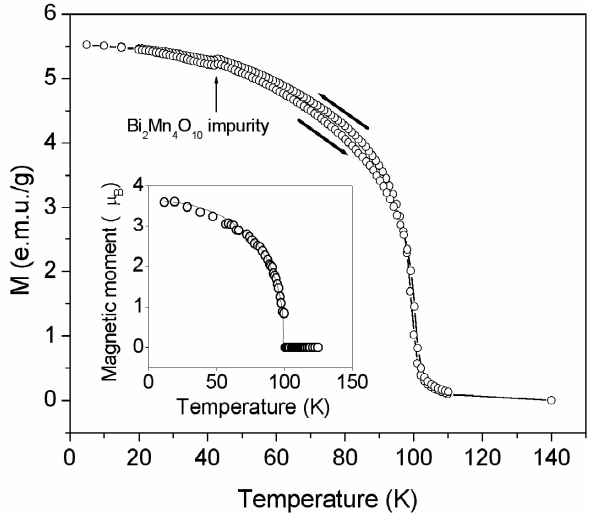}
\caption{\label{Fig_1}\textbf{Main panel}: Magnetization vs. temperature for BiMnO$_3$, collected on both heating and cooling in a 100 Oe magnetic field. The anomaly at 42 K (arrow) is due to a small amount of antiferromagnetic Bi$_2$Mn$_4$O$_{10}$, which does not show hysteretic behavior. \textbf{Inset}: Mn magnetic moment versus temperature, as determined from neutron diffraction data (see text).}
\end{figure}

Polycrystalline samples of BiMnO$_3$ were synthesized by solid state reaction of Bi$_2$O$_3$ (Aldrich 99.99 \%) and Mn$_2$O$_3$ (Aldrich 99.999 \%) at 40 Kbar and 600 $^{\circ}$C for 3 hours, using a high-pressure multi anvil apparatus. Three samples were synthesized at the same experimental conditions in order to obtain a suitable quantity of powder for neutron diffraction experiment ($\sim$ 2g).
Magnetization curves were measured in the temperature range 5-20 K in steps of 5 K and from 20 to 110K in steps of 1K with a superconducting quantum device (SQUID) magnetometer (Quantum Design). The magnetization curves, measured in a 100 Oe magnetic field both on cooling and in heating mode (Fig. \ref{Fig_1}),  evidenced the presence of  a small thermal hysteresis (T$_C$ (heating)= 100.26 K,  T$_C$ (cooling) = 99.05), consistent with the magnetic transition having a slight first-order character.

Neutron powder diffraction data were collected at room temperature (RT) using the high resolution powder diffractometer (HRPD) and as a function of temperature using the general materials diffractometer (GEM) at the ISIS facility. Data were collected on warming from 10 to 293 K in steps ranging from 1 to 10 K using a helium cryostat. Different collection times (20 min- 6 h) were employed to obtain higher statistics at key temperatures. Additional measurements were performed on GEM at a fixed temperature (95K) using a cryomagnet to vary the magnetic field from 0 to 10 T,  in steps of 0.5 T.  Rietveld refinements of the structure (nuclear above T$_C $ and nuclear and magnetic below T$_C $) were performed at all temperatures in both $C2$ and $C2/c$ space groups, using the program GSAS \cite{Larson}. In the course of the refinements, the magnetic moments on the inequivalent Mn sites were constrained to be equal, and to lie along the high-symmetry $y$ direction.

The structural parameters as refined on the RT HRPD data  in $C2$ and $C2/c$ are reported in Tab. I of the Supplementary Material.   Both the quality of the refinements and the actual structural parameters are extremely similar in the two cases, resulting in similar bond distances and comparable formal oxidation states for Bi and Mn.  In spite of the larger number of parameters, the $R$ factors obtained in $C2$ are slightly higher than those in $C2/c$. In addition, the estimated standard deviations of the atomic coordinates are one order of magnitude larger in $C2$, indicating the presence of significant correlations.  We therefore conclude that the \emph{centrosymmetric} space group $C2/c$ is the appropriate one to describe the structure.  The same conclusion applies to the structural and magnetic refinements of the 10K data collected on GEM.  A comparison of the RT and 10 K structures (Tab. II  and III of the Supplementary Material) indicates that the  distribution of shorter-basal and longer-apical bonds is essentially the same, confirming that the orbital ordering pattern is unchanged on cooling.  The refined magnetic moment per Mn site as a function of temperature is reported in the inset of Fig. \ref{Fig_1}, showing a good agreement with bulk magnetization measurements.

\begin{figure}
\includegraphics[scale=0.40]{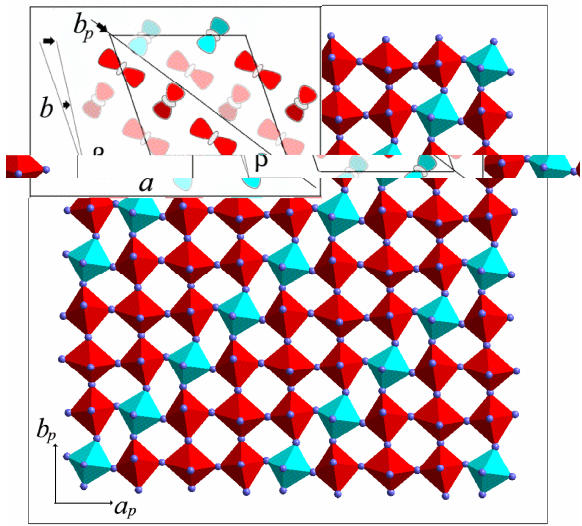}
\caption{\label{Fig_2}\textbf{color online}.  Orbital ordering pattern in BiMnO$_3$. \textbf{Main Panel}:  A slice cut perpendicular to the pseudo-cubic c$_p$ axis, showing MnO$_6$ octahedra with in-plane (red) and out-of-plane (light blue) long Mn-O bonds, corresponding to occupied d$_{z^2-r^2}$ e$_g$ orbitals.  \textbf{Inset}:  the same structure shown on the monoclinic axes.  The presentation emphasized the occupied orbitals.}
\end{figure}

The pattern of orbital ordering giving rise to ferromagnetism is best visualized by referring it to the pseudo-cubic perovskite axes (Fig. \ref{Fig_2}), which also make it very easy to compare BiMnO$_3$ with other orbitally-ordered perovskites such as LaMnO$_3$.  The appropriate cell is actually triclinic, it encompasses $4 \times 4 \times 4$ primitive perovskite units and is related to the monoclinic cell by the covariant transformation

\begin{equation}\label{Eq:_covariant}
\left[\textbf{a}_p, \textbf{b}_p, \textbf{c}_p\right]= \left[\textbf{a}_m, \textbf{b}_m, \textbf{c}_m\right]\begin{scriptsize} \left[ \begin {array}{ccc} 1&1&1\\\noalign{\medskip}2&0&-2
\\\noalign{\medskip}1&-1&1\end {array} \right]\end{scriptsize}
\end{equation}

with $a_p = c_p = 4 \times 3.92$ \AA, $b_p = 4 \times 3.98$ \AA,  $\alpha_p =  \gamma_p = 91.4^{\circ}$ and  $\beta_p = 90.8^{\circ}$.  Fig. \ref{Fig_2} shows a section of this structure perpendicular to the $c_p$ axis.  Subsequent layers are staggered by one perovskite unit along the  $a_p$ direction.  One should also bear in mind that $a_p-b_p$ and $c_p-b_p$ sections are identical, since they are related by the mirror symmetry in $C2/c$.  In Fig. \ref{Fig_2}, dark red octahedra have their partially filled orbital (long bonds) \emph{within} the projection plane, whereas light blue octahedra have it \emph{perpendicular} to it.  In the case of LaMnO$_3$, the same section (through the $b$ axis in the $Pnma$ setting) would only show dark red octahedra with identical arrangement.  Based on the well-known Goodenough-Kanamori rules, it is therefore easy to understand that the dark red "stripes" carry a uniform FM interaction as in the equivalent LaMnO$_3$ layers.  The magnetic interactions along the $b_p$ direction are also uniformly FM. Interactions around "light blue" octahedra are partially frustrated, 4 FM and 2 antiferromagnetic (AFM), but the FM interactions, which are also expected to be larger in magnitude, clearly dominate, giving rise to an overall FM structure.

\begin{figure}[h!]
\includegraphics[scale=0.42]{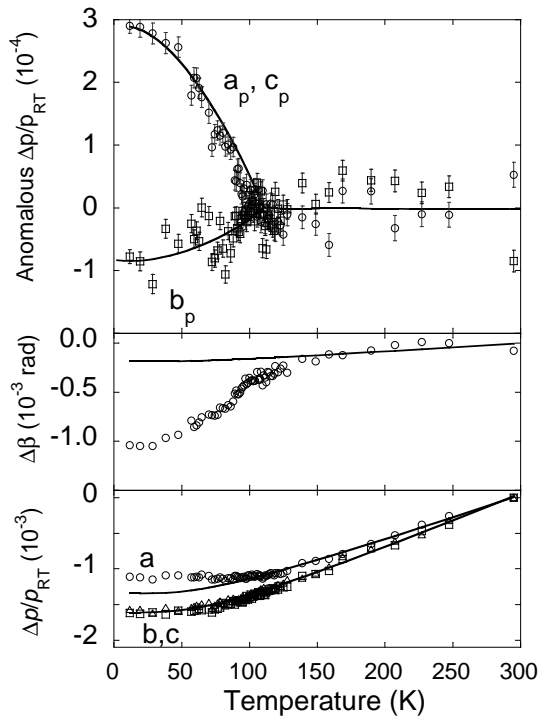}
\caption{\label{Fig_3} Magneto-elastic effect in BiMnO$_3$.  \textbf{Bottom}: Relative change ($\frac{\Delta p (T)}{p(300 K)})$ of the monoclinic lattice parameters $p=a$, $b$ and $c$ and \textbf{Middle}: change of the monoclinic $\beta$ angle vs. temperature. The parameters were determined from Rietveld refinements based on neutron diffraction data (symbols). The lines represent the calculated thermal expansion due to phonons, based on a simple Einstein fit to the high-temperature data with T$_E$ = 200 K(see text).   \textbf{Top}, excess thermal evolution (i.e., with the phonon contribution subtracted off) of the pseudo-cubic lattice parameters.  Lines are guides to the eye.  If not shown, the error bars are smaller than the symbols.}
\end{figure}

\begin{figure}[ht!]
\includegraphics[scale=0.42]{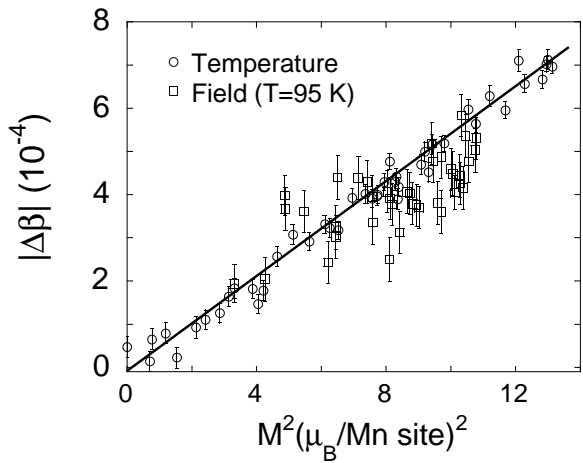}
\caption{\label{Fig_4}Change of the monoclinic $\beta$  angle as a function of the \emph{square} of the ordered magnetic moment on the Mn site, as obtained by varying the temperature in zero field  (\textbf{circles}) and by varying the applied magnetic field in the range 0-10 T at a constant temperature of 95 K (\textbf{squares}).}
\end{figure}

The bottom and middle panels of Fig. \ref{Fig_3} display the evolution of the monoclinic lattice parameters as a function of temperature.  The dimensionless normalization ($\begin{scriptsize}(p(T)-p(RT))/p(RT))\end{scriptsize}$ for $p=a$, $b$ and $c$ and $\begin{scriptsize}\beta(T)-\beta(RT)\end{scriptsize}$ for the monoclinic angle in radiants) is appropriate for the calculation of the strain tensor (see below).  There is a clear anomaly with onset at T$_C$ for the monoclinic angle and, to a lesser extent, for $a$, while $b$ and $c$ are hardly affected by the transition. The solid lines represent fits to the data above the transition using a simple Einstein model of the phonon thermal expansion, with T$_E$=200 K (the results are affected only slightly by the exact value of T$_E$).  This enabled us to extract the "anomalous" component of the lattice distortion due to magnetoelasticity (see below).  The top panel of Fig. \ref{Fig_3} shows this anomalous distortion after projection onto the pseudo-cubic axes (the effect of the transition on the triclinic angles is very small).  This presentation clarifies the origin of the magnetoelastic effect: the lattice \emph{shortens} along the unfrustrated direction $b_p$ (Fig.2, inset), and it \emph{lengthens} to a greater extent along the two other partially frustrated directions.  The overall effect on the unit cell volume is a slight expansion on cooling (negative magnetostriction, not shown).  The magnetoelastic origin of these lattice anomalies finds further support in the data we collected as a function of applied magnetic field at fixed temperature (T = 95K, just below Tc).  Fig. \ref{Fig_4} demonstrates that the anomalous variation of the $\beta$ angle (with the small phonon component subtracted) is proportional to the \emph{square} of the ordered magnetic moment, regardless of whether the moment is induced by temperature or field.

In order to relate these changes to the magnetodielectric anomalies observed by Kimura \textit{et al.} \cite{Kimura}, we need to consider an appropriate form of the Landau free energy including the lattice strain.  Since the structure remains monoclinic below T$_C$, the strain does not break the symmetry, and all the elements of the strain tensor are in principle allowed to contribute independently to the free energy.  However, for simplicity, we will only consider the \emph{scalar} strain e$_{sc}$, essentially the quadrature sum of the elements of the strain tensor.  In the monoclinic case, it can be shown \cite{Salje} that only the e$_{33}$ and e$_{31}$ tensor elements contain $\Delta \beta$, and that, for the relevant values of the parameters and ignoring the small $a$ anomaly, $\textrm{e}_{sc}\approx \kappa \beta$, with $\kappa \simeq 1$.  Therefore, Fig. \ref{Fig_4} provides a \emph{quantitative} estimate of the magnetoelastic strain.  A similar result can be obtained by considering the \emph{diagonal} elements of the strain in the triclinic setting (Fig. \ref{Fig_3}, top).  The magnetoelastic strain is one order of magniture \emph{larger} than the one reported by Kimura on the basis of dilatometric measurements - a discrepancy most likely due to the powder averaging in the latter.  In the Landau free energy expansion, the terms responsible for the magnetodielectric and magnetoelastic coupling are $\frac{\zeta}{2} M^2P^2$ and $\xi e_{sc}M^2+\frac{c}{2}e_{sc}^2$, respectively, where $M$ is the magnetization, $P$ is the polarization and $\zeta$,$\xi$ and $c$ are phenomenological constants.  By minimizing the free energy, one can show that, near the transition and for large values of the dielectric constant $\epsilon$,

\begin{eqnarray}
e_{sc}&=&-\frac{\xi}{c}M^2\\
\frac{\Delta \epsilon}{\epsilon_0}& \simeq &\epsilon_0 \zeta M^2 \nonumber
\end{eqnarray}

In other words, both the dielectric anomaly and the magnetoelastic strain are proportional to the square of the magnetization, and therefore to each other.  However, since the coupling constants involved are different, there is no reason why the proportionality constant should be of the order 1 (it is in fact $\sim$ 20).  Here, we have clearly shown the presence of a structural rearrangement upon magnetic ordering in BiMnO$_3$, as evidenced by very significant lattice anomaly.  However, the most likely microscopic origin of the observed dielectric anomalies is not the change in lattice metrics itself, but rather a related subtle change in the environment of the highly polarizable Bi$^{3+}$ ion.  Our data both above and below the magnetic transition confirm that, within our accuracy, the structure of BiMnO$_3$ retains a center of inversion, and is therefore very unlikely to be ferroelectric.

\end{document}